\documentclass[11pt,a4paper]{article}
\usepackage{jcappub}
\usepackage{bm}

\newcommand{\fnl}{\ensuremath{f_{\rm NL}}}
\newcommand{\bphi}{\ensuremath{b_\Phi}}
\newcommand{\s}{\ensuremath{\bm{s}}}
\newcommand{\nsel}{\ensuremath{n_{\rm sel}}}

\title{Estimating non-gaussian bias using counts of tracers}
\author[a]{Neal Dalal,}
\author[b,c,a]{Will J. Percival}
\affiliation[a]{Perimeter Institute for Theoretical Physics\\31 Caroline St. N, Waterloo, Ontario N2L 2Y5, Canada}
\affiliation[b]{Waterloo Centre for Astrophysics, University of Waterloo\\ Waterloo, Ontario N2L 3G1, Canada}
\affiliation[c]{Department of Physics and Astronomy, University of Waterloo\\ Waterloo, Ontario N2L 3G1, Canada}
\emailAdd{ndalal@perimeterinstitute.ca}
\abstract{
Local-type primordial non-gaussianity generates a distinctive term in the clustering of tracers of large-scale structure, behaving as $k^{-2}$ at small wavenumbers $k$. In order to use this signal in a sample of galaxies to measure the amplitude of primordial non-gaussianity, \fnl, we need to independently determine the degenerate bias coefficient, \bphi, which quantifies the  logarithmic response of the galaxy number density to a change in amplitude of the matter clustering. We study whether \bphi\ may be estimated from the observed evolution of the number density of galaxies as a function of redshift.  Using cosmological N-body simulations, we find that \bphi\ may be estimated reasonably well for dark matter halos across the range of redshifts and halo masses used by large-scale structure surveys aimed at measuring \fnl.  This includes non-gaussian secondary bias (or assembly bias) in halo concentration, which has previously been found to be quite large in amplitude. For an observed survey of galaxies, we additionally need to consider the selection function of the sample, which can introduce redshift dependence via cuts on apparent magnitude and colour.  These effects of the selection function can be mitigated by further cutting the sample using k-corrected magnitudes and colours, to retain only those galaxies that would pass the targeting criteria for all redshifts within the interval considered.}
\notoc
\toccontinuoustrue

\begin{document}

\maketitle

\section{Introduction}

The amount and form of non-gaussianity of primordial density and curvature perturbations provide a powerful probe of the fields present during cosmic inflation and their interactions \cite{Maldacena2003,Arkani2015}.  We can probe this non-gaussianity through observations of higher-order correlations of the CMB anisotropies \cite{Komatsu2003,Planck2020}, or of the matter field in the low-redshift universe \cite{Cabass2022,Cabass2025,Damico2025}.  An initially surprising but now well-known result is that we can \emph{also} probe non-gaussianity using the 2-point correlations of tracers of large-scale structure, like galaxies or quasars \cite{Dalal2008}.  The reason is that the clustering bias of these tracers is sensitive to the squeezed limit of the primordial bispectrum of the gravitational potential \cite{Assassi2015,Desjacques2018}.  A well-studied example is so-called ``local'' primordial non-gaussianity, where the non-gaussian potential $\Phi$ can be expressed in terms of a Gaussian random field $\Phi_G$ as 
\begin{equation} \label{eq:localng}
\Phi(\bm{x}) = \Phi_G(\bm{x}) 
    + \fnl[\Phi_G^2(\bm{x})-\langle\Phi_G^2\rangle],
\end{equation}
where the parameter \fnl\ characterizes the amplitude of primordial non-gaussianity \cite{Komatsu2003}.  A significant detection of nonzero \fnl\ would exclude single-field inflation, and therefore considerable efforts have been devoted to constraining this cosmological parameter \cite{Dore2014,Ferraro2022}. 
Local non-gaussianity couples long-wavelength perturbations to small-scale perturbations, in a particularly simple way:
\begin{equation}
P_{\rm local}(k_s) \approx P(k_s)\,[1 + 4\fnl\Phi(\bm{k}_l)].
\end{equation}
That is, a long-wavelength mode $\Phi(\bm{k}_l)$ rescales the entire small-scale power spectrum, with a change proportional to $\fnl\Phi(\bm{k}_l)$.
This modulation of the small-scale power spectrum necessarily gives a modulation of the abundance $n$ of tracers, since the abundance depends on the small-scale power spectrum.  Since the change in the small-scale power spectrum is $\delta P \approx 4\fnl \Phi(\bm{k}_l) P$, then $\delta n \approx (\partial n/\partial P) \delta P \approx 4\fnl \Phi(\bm{k}_l) P \,\partial n/\partial P$, and $\delta n/{\bar n} \approx 4\fnl \Phi(\bm{k}_l) \partial\log\bar n/\partial\log P = 2 \fnl \Phi(\bm{k}_l) \partial\log\bar n/\partial\log\sigma_8$, using $P \propto \sigma_8^2$ \cite{Biagetti2019}. This is commonly written in the literature as
\begin{equation}
\delta(\bm{k}) = \bphi \fnl \Phi(\bm{k}),
\end{equation}
where $\bphi = 2\partial\log\bar n/\partial\log\sigma_8$.  This fluctuation in the tracer density adds to all the other fluctuations on large scales, such as $b_1 \delta_m(\bm{k})$.  Since the primordial $\Phi(\bm{k}) \propto \delta_m(\bm{k})/k^2$ on large scales, we can see that this $\Phi$ term has the appearance of a correction to $b_1$ that scales like $\fnl/k^2$, and has thus been called a scale-dependent bias.

Detecting this scale-dependent bias would thus detect local-type primordial non-gaussianity.  However, since \fnl\ is always paired with \bphi, a measurement of scale-dependent bias measures the product \fnl\bphi.  To determine \fnl, we need to independently know the value of \bphi.  
Early work on simulations with local PNG found that for dark matter halos selected on virial mass, there is a simple relationship between \bphi\ and the linear bias $b_1$ \cite{Dalal2008}.  Since $b_1$ can be accurately measured from clustering on scales much smaller than the horizon, this $b_1-\bphi$ relationship can therefore be used to predict \bphi, for halos selected on virial mass.  Subsequent work quickly showed, however, that halos selected on quantities besides virial mass can deviate significantly from the $b_1-\bphi$ relation for halo virial mass \cite{Slosar2008,Reid2010}.  This suggests that 
for an arbitrary tracer population (e.g., some type of galaxy or quasar) which populates halos in a potentially complicated way that depends on more than virial mass, \bphi\ cannot be predicted, and instead must be treated as a nuisance parameter in the same way that $b_1$ is considered as a nuisance parameter.  Marginalizing over arbitrary \bphi\ would preclude meaningful constraints on \fnl\ from being drawn, and therefore several works have studied whether priors may be placed on \bphi, for example using hydrodynamic simulations of galaxy formation \cite{Barreira2020,Fondi2024}. 

In this brief note, we consider an alternative approach, in which we investigate whether \bphi\ may be estimated directly from observed data.  In the next section, we outline the basic idea, and then in subsequent sections we apply this method to cosmological N-body simulations and consider how we can mitigate selection effects in an actual galaxy survey.

\section{Estimating \bphi\ from data}

Consider a tracer population observed over a range of redshift with expected comoving number density ${\bar n}(z)$.  In general, ${\bar n}(z)$ will depend on many quantities, including the selection function, and the background cosmology via $\sigma_8$, $\Omega_m$, etc.  The redshift derivative of the number density can be expanded
\begin{equation} \label{eq:dndz}
\frac{d\log\bar n}{dz} = \frac{\partial\log\bar n}{\partial\log\sigma_8} \frac{d\log\sigma_8}{dz} + 
\frac{\partial\log\bar n}{\partial\Omega_m} \frac{d\Omega_m}{dz}
+ \ldots + \frac{\partial\log\bar n}{\partial z}.
\end{equation}
In this expression, quantities like $\sigma_8$, $\Omega_m$, etc., all evolve with $z$, and the final term on the right-hand side accounts for any explicit redshift dependence of the number density at fixed $\sigma_8$, $\Omega_m$, etc.  

The left-hand side of eq.\ \eqref{eq:dndz} is directly observable, and the first term on the right-hand side involves \bphi.  If the other terms on the RHS could be measured, or a sample selected such that they could be neglected compared to the first term, we could then estimate \bphi\ using 
\begin{equation} \label{eq:estimatebphi}
\frac{\partial\log\bar n}{\partial\log\sigma_8} \approx 
\frac{d\log\bar n/dz}{d\log\sigma_8/dz},
\end{equation}
i.e.\ the evolution observed across redshift could be used to determine \bphi.  In practice, this could be difficult to achieve exactly, given the evolution of other cosmological parameters and sample selection.  However, there are scenarios where this should be a good approximation.
One example where the other terms on the right-hand side of eq.\ \eqref{eq:dndz} may be neglected is for cold dark matter halos selected on virial mass, because the abundance of such halos is nearly `universal' in form \cite{Tinker2008}, with a mass function that may approximately be expressed as
\begin{equation} \label{eq:universal}
\frac{d\bar n}{d\log M} = \frac{\bar\rho_{m0}}{M}
\left|\frac{d\log\sigma}{d\log M}\right|f(\sigma),    
\end{equation}
for some function $f$ that depends only on 
$\sigma^2$, the variance of the linear overdensity smoothed on mass scale $M$.  For universal mass functions, all of the time dependence of $\bar n$ arises from the time dependence of $\sigma \propto \sigma_8$, meaning that the other terms on the right-hand side of eq.\ \eqref{eq:dndz} vanish, and therefore eq.\ \eqref{eq:estimatebphi} should provide a good estimate of \bphi.  In the next section, we apply eq.\ \eqref{eq:estimatebphi} to halos selected on virial mass in N-body simulations, and unsurprisingly we find that this provides a good estimate of \bphi.

For more general tracers, whose abundance is not universal in the form of eq.\ \eqref{eq:universal}, $\bar n$ can depend on other quantities like $\Omega_m(z)$, making eq.\ \eqref{eq:estimatebphi} invalid.  Even for tracers whose intrinsic abundance is universal, the observational selection function may have explicit redshift dependence, similarly invalidating this expression.

Nevertheless, because the numerator of eq.\ \eqref{eq:estimatebphi} is directly observable, and the denominator is known (as long as the background cosmological parameters like $\Omega_{m0}$ and $w_{\rm DE}$ are known), using this expression is attractive as an empirical way to roughly estimate \bphi.  In the next sections, we consider the  conditions under which this approximation may be employed.

\section{N-body halos}

\begin{figure}
    \centering
    \includegraphics[width=0.48\linewidth]{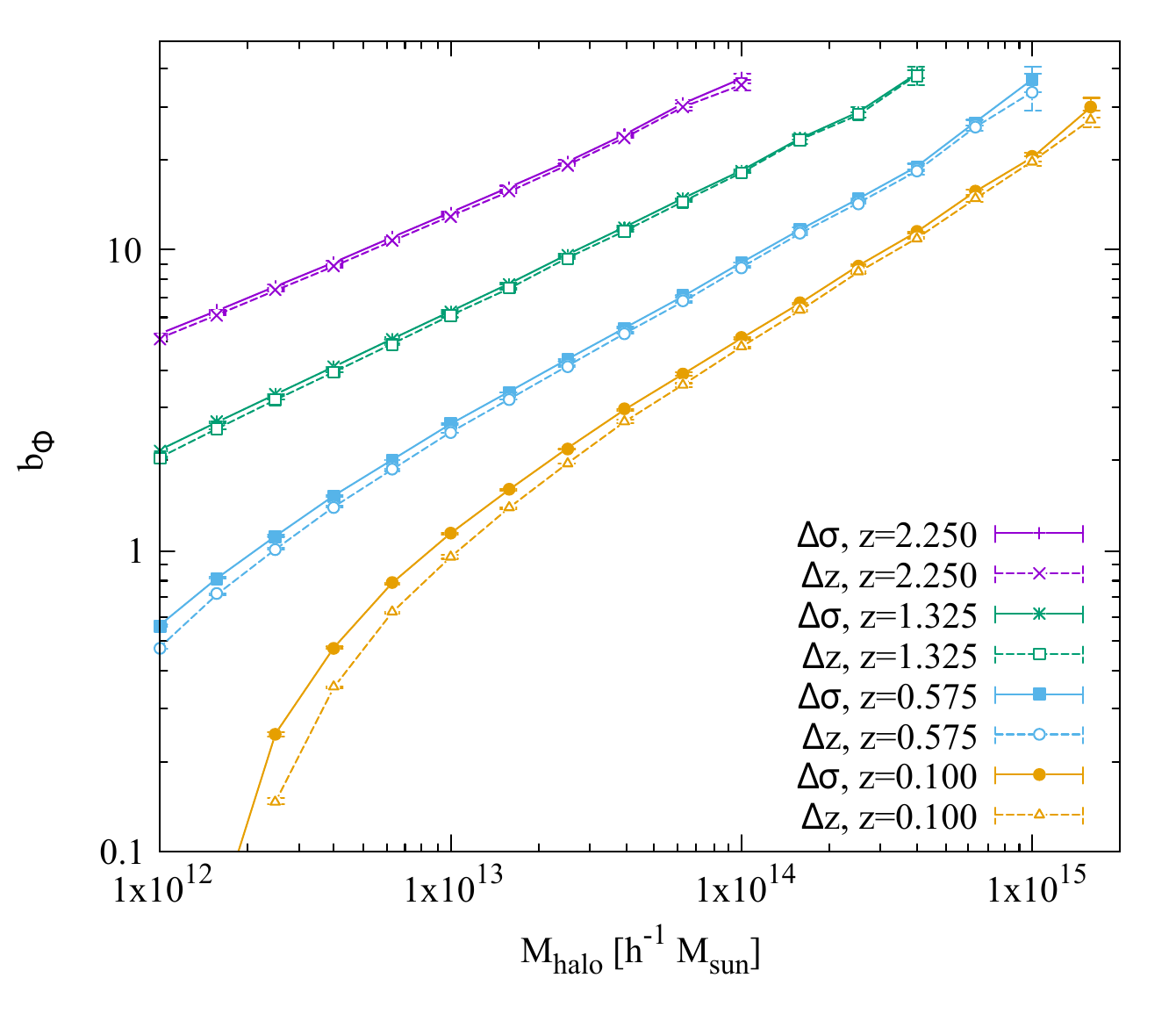} 
    \hfill
    \includegraphics[width=0.48\linewidth]{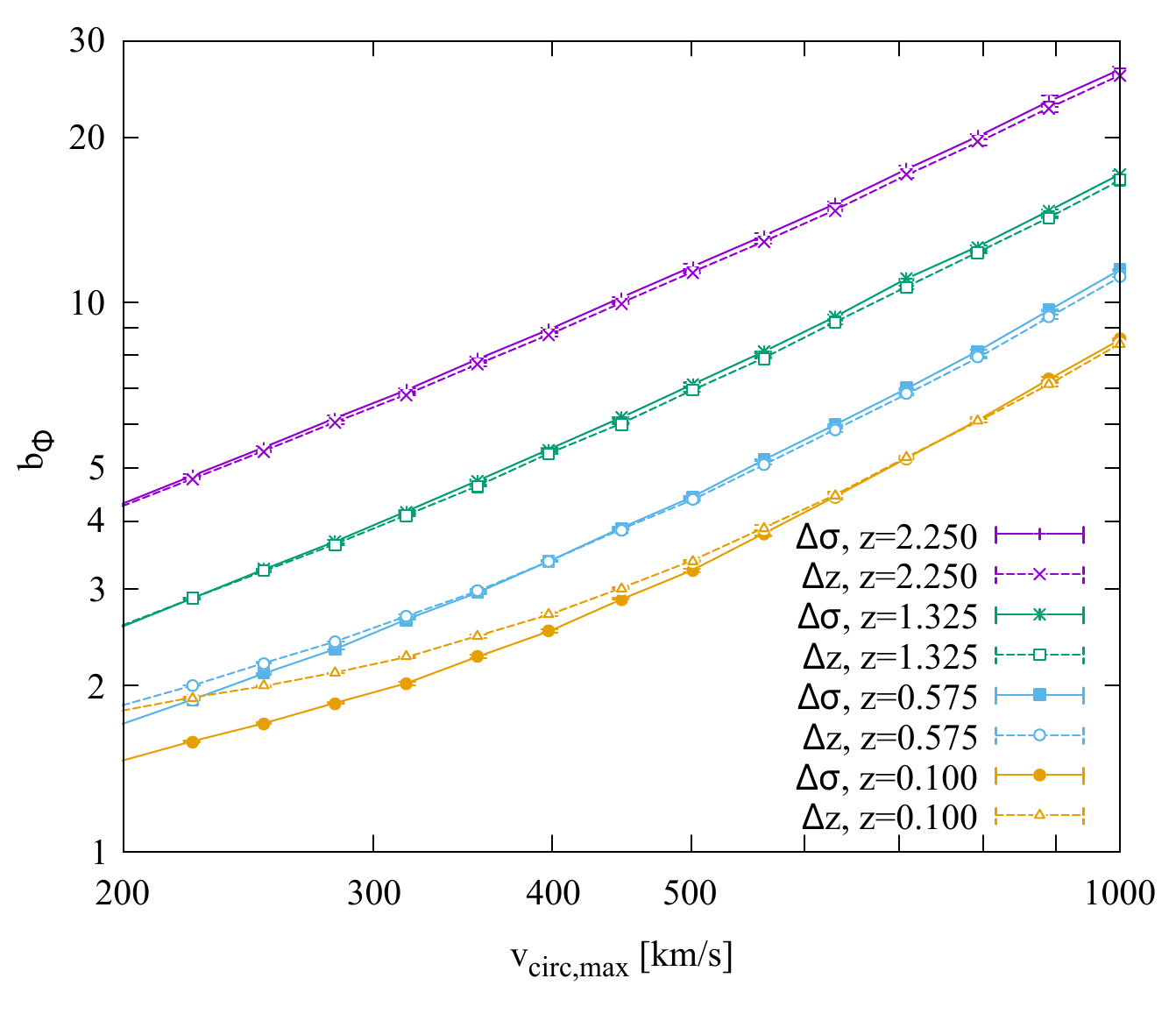}
    \caption{Comparison of \bphi\ measured directly using N-body simulations with different $\sigma_8$ ($\Delta\sigma$, solid lines), vs.\ \bphi\ estimated using snapshots at different redshifts from the same simulation ($\Delta z$, dashed lines).  {\bf Left}: In this panel we show \bphi\ as a function of halo mass for halos selected solely on virial mass, at various redshifts indicated by the different colours.  The agreement between the solid and dashed lines illustrates that eq.\ \eqref{eq:estimatebphi} is a good approximation for halos with a nearly universal mass function.
    {\bf Right}: Similar to the left panel, but now for halos selected on peak circular velocity $v_{\rm circ,max}$; see main text for more detail.  We again find good agreement for a range of redshifts, with the worst disagreement found at $z\approx 0$.} 
    \label{fig:nbody}
\end{figure}

In this section, we use N-body simulations to study when eq.\ \eqref{eq:estimatebphi} may be used to provide a reasonable estimate of \bphi.  We use the publicly available {\tt CompaSO} halo catalogs \cite{Hadzhiyska2021} from the Abacus Summit simulations \cite{Maksimova2021,Garrison2021}.  This simulation suite provides multiple snapshots at different redshifts, allowing us to measure the redshift differences used in the right-hand side of eq.\ \eqref{eq:estimatebphi}.  It also provides simulations with identical cosmological parameters and identical random initial seeds, except for different values of $\sigma_8$, allowing us to measure \bphi\ exactly, avoiding the complications of inference based on evolution.  Specifically, we use base simulation sets {\tt c000} and {\tt c004}, which both simulate $(2\, h^{-1} {\rm Gpc})^3$ volumes using $6912^3$ particles and Planck2018 parameters, except for $\sigma_8=0.75$ used in {\tt c004}.  We used 4 boxes for these two sets, namely {\tt ph000, ph001, ph002, ph003}.  From these simulations, we examined halo catalogs for redshifts $z=0.100,\, 0.575,\, 1.325,\, 2.250$. 
    
First we consider \bphi\ for halos selected solely by their mass. Here, eq.\ \eqref{eq:estimatebphi} provides a good approximation to \bphi, as shown in the left panel of figure \ref{fig:nbody}.  
As discussed in the previous section, the agreement with eq.\ \eqref{eq:estimatebphi} makes sense because for this halo sample, the other terms in the right-hand side in eq.\ \eqref{eq:dndz} are expected to be negligible compared to the \bphi\ term.  

If we select halos using properties other than halo mass, then we can no longer use universality to neglect the other terms in eq.\ \eqref{eq:dndz}, making it unclear whether eq.\ \eqref{eq:estimatebphi} will continue to provide a reasonable approximation.  For example, consider counting halos selected on their circular velocity.  If we compare counts of halos with the same peak circular velocity $v_{\rm circ,max}$ evaluated at different redshifts, we would find that eq.\ \eqref{eq:estimatebphi} provides a poor estimate of \bphi.  Part of the reason for this, however, is that the circular velocity in a halo is expected to depend on formation time, both because of the expansion of the universe, but also because the virial overdensity $\Delta_{\rm vir}$ evolves when $\Omega_m \neq 1$.  We can remove some of that expected time dependence by comparing halos of fixed $v/(\Delta_{\rm vir}^{1/6} (1+z)^{1/2})$ 
at different redshifts when evaluating the redshift derivative in eq.\ \eqref{eq:estimatebphi}, instead of comparing halos of fixed $v$. The right panel of figure \ref{fig:nbody} shows the resulting estimate of \bphi, compared to the actual value of \bphi.  Making this small correction results once again in a good estimate for \bphi\ across a range of redshifts and masses, with the worst discrepancies ($\sim 40\%$) occurring near $z\sim 0$.  

The reasonably good agreement shown in the right panel of figure \ref{fig:nbody} is nontrivial, unlike the left panel, because the halo velocity function $d\bar n/d\log v_{\rm circ,max}$ is not expected to be nearly universal in the same way that the halo mass function $d\bar n/d\log M$ is.  This is because $v_{\rm circ,max}$ depends not only on halo mass $M$, but also on halo concentration.  Note that removing the time-dependent virial velocity scaling  $\Delta_{\rm vir}^{1/6} (1+z)^{1/2}$ does not affect the dependence of $v_{\rm circ,max}$ on concentration.
Previous work has shown that \bphi\ is a function of secondary halo properties like concentration, and moreover this non-gaussian secondary bias in \bphi\ behaves differently than gaussian secondary bias in $b_1$, leading to large discrepancies from the standard $b_1-\bphi$ relation of mass-selected halos \cite{Lazeyras2023}.  The good agreement in \bphi\ for velocity-selected halos suggests that non-gaussian secondary bias from parameters like concentration might not ruin the approximate estimate of \bphi.  Stated another way, figure \ref{fig:nbody} suggests that eq.\ \eqref{eq:estimatebphi} may provide a good estimate of non-gaussian bias from secondary parameters like concentration.

Naively, there are good reasons to doubt that eq.\ \eqref{eq:estimatebphi} should provide a reasonable estimate of the concentration dependence of \bphi.  
Physically, we can expect the process of halo collapse and formation to differ somewhat at low redshift (when $\Omega_m<1$) compared to high redshift (when $\Omega_m \approx 1$).  One example of this is the redshift dependence of the splashback feature in halo density profiles, which arises due to the effects of dark energy on the turnaround, infall and orbits of particles entering halos at low redshifts \cite{Adhikari2014}.  Using similar reasoning, we can expect halo internal structures also to depend on the relative densities of dark energy and dark matter, meaning that when we select halos based on their structural parameters like concentration, the resulting abundance should depend on $\Omega_m(z)$, giving nonzero $\partial \log\bar n/\partial\Omega_m$ in eq.\ \eqref{eq:dndz} and therefore potentially significant errors in eq.\ \eqref{eq:estimatebphi}.

\begin{figure}
    \centering
    \includegraphics[width=0.95\linewidth]{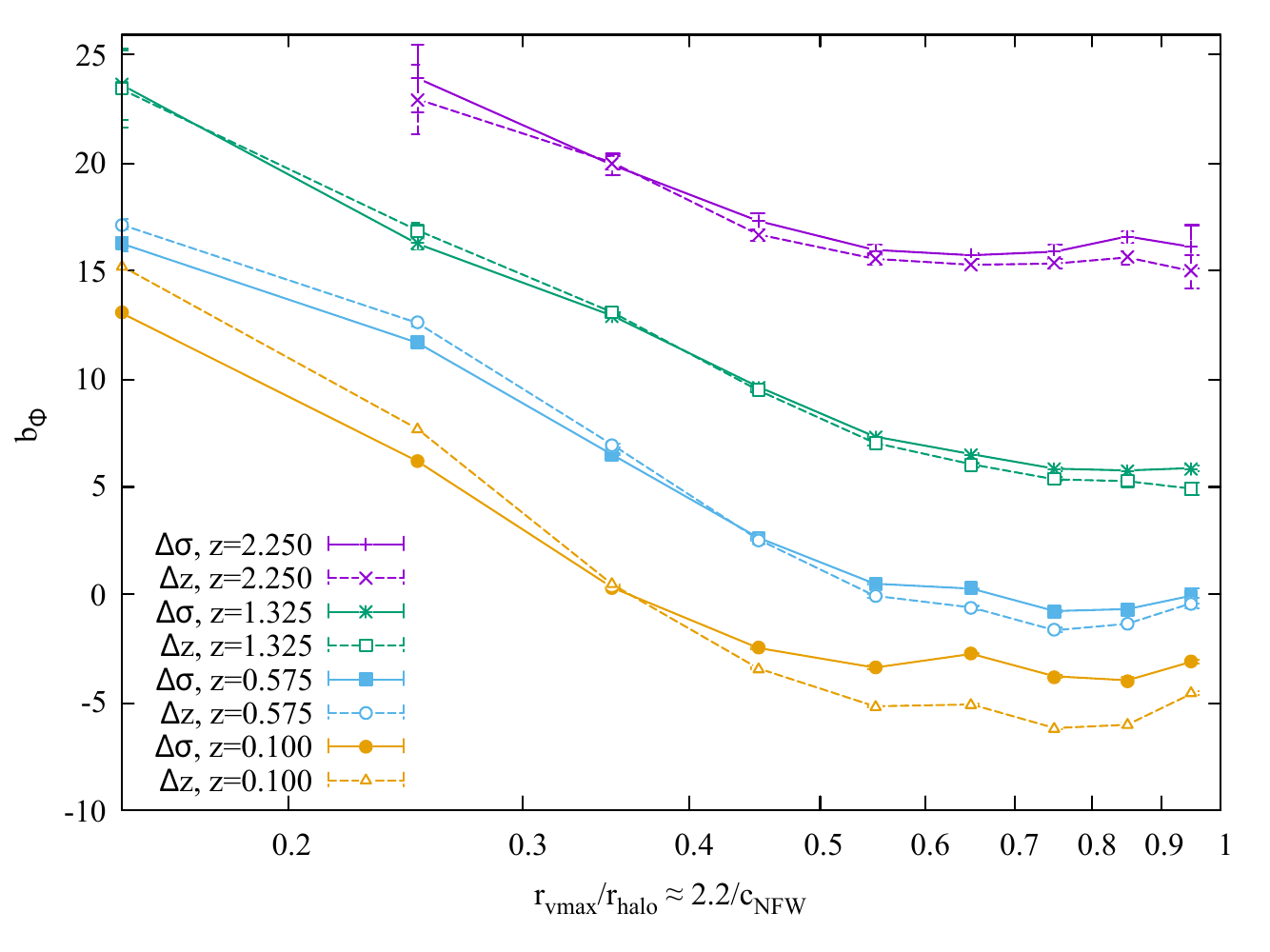}
    \caption{Non-gaussian bias \bphi\ 
for halos of different concentration, to illustrate non-gaussian secondary bias (or assembly bias).  We select halos with 6000-10000 particles, corresponding to the mass range $1.3-2.1\times10^{13} h^{-1}M_\odot$, and then plot \bphi\ as a function of the ratio $r_{v\rm max}/r_{\rm halo}$, where $r_{v\rm max}$ is the radius where the circular velocity peaks.  Even at fixed halo mass, \bphi\ depends significantly on concentration, but this secondary bias appears to be approximated well by eq.\ \eqref{eq:estimatebphi} for a range of redshifts, with the worst disagreement found at $z\approx 0$. Lines and symbols are as in Fig.~\ref{fig:nbody}.} 
\label{fig:secondarybias}
\end{figure}

To check this, in figure \ref{fig:secondarybias} we show the dependence of \bphi\ on halo concentration.  We select halos in the mass range $1.3-2.1\times 10^{13} h^{-1} M_\odot$, as an example typical for the DESI LRG sample \cite{Yuan2024}.  We measure halo concentration using the ratio $r_{v \rm max}/r_{\rm halo}$ reported in the {\tt CompaSO} catalogs \cite{Hadzhiyska2021}, which should be approximately equal to $2.2/c_{\rm NFW}$ for a NFW profile \cite{Navarro1997}.  As suggested by figure \ref{fig:nbody}, we see that eq.\ \eqref{eq:estimatebphi} does indeed provide a reasonable estimate of the concentration dependence of \bphi, with the worst agreement found near $z\sim 0$.  Since there is little comoving volume at $z\sim 0$, figure \ref{fig:secondarybias} suggests that eq.\ \eqref{eq:estimatebphi} may provide a useful estimate of \bphi\ over the redshift range used to probe \fnl.

This good agreement perhaps makes sense at large redshift, e.g.\ $z>1$, where $\Omega_m \approx1$, but it may be somewhat surprising that eq.\ \eqref{eq:estimatebphi} remains accurate even at $z\sim 0.5$, since $d\Omega_m/dz$ is nearly as large at $z=0.575$ as it is at $z=0.1$.  One possible explanation for this good agreement down to $z\sim 0.5$ is that halo concentration is sensitive to halo interiors, which were assembled earlier than halo outskirts \cite{Wechsler2002,Dalal2010}.  Indeed, we do see in figure \ref{fig:secondarybias} that the worst agreement is found for low-concentration halos, where a greater proportion of mass accreted recently.  This suggests that eq.\ \eqref{eq:estimatebphi} may become a poor approximation earlier than shown in figure \ref{fig:secondarybias} for secondary parameters more sensitive to halo outskirts.

\section{Galaxy Surveys} \label{sec:k-correction}

The results shown in Figs.\ \ref{fig:nbody} and \ref{fig:secondarybias} are encouraging for surveys that aim to probe \fnl, as they suggest that it may be possible to estimate \bphi\ using eq.\ \eqref{eq:estimatebphi} over a broad redshift range for a subsample of tracers that behave similarly to dark matter halos. Although we cannot easily select individual galaxies based on their halo properties, upcoming surveys like SPHEREx \cite{Dore2014}, Euclid \cite{Ballardini2024}, or LSST \cite{Ivezic2019} may allow measurement of accurate stellar masses of detected galaxies.  Since the stellar mass - halo mass relation (SHMR) evolves only mildly with redshift  \cite{Shuntov2022}, this may allow us to apply eq.\ \eqref{eq:estimatebphi} directly to samples selected on stellar mass.  Evolution in the SHMR could produce errors in the inferred \bphi.  Schematically, if we write the tracer density as $\bar n = \bar n_h\,\bar N_t$, where $\bar n_h$ is the halo abundance and $\bar N_t$ is the mean number of tracers per halo, then $\partial_z\log\bar n = \partial_z\log\bar n_h + \partial_z\log\bar N_t$.  To avoid large errors in the estimated \bphi, we therefore require the change in $\log\bar N_t$ to be small compared to the change in $\log\bar n_h$ over the observed redshift interval.  Since the tracer halo occupation distribution (HOD) may be precisely measured using measurements of small-scale clustering \cite{Yuan2024}, it should be possible to check whether this condition is satisfied.

Besides selecting on stellar mass, we can also consider more generally selecting samples based on other rest-frame properties, independently of the galaxy's redshift.  However, this type of selection is not typically employed by most large-scale structure surveys.  Rather, selection cuts are more commonly based on each galaxy's observed (redshifted) spectrum, instead of the galaxy's rest-frame spectrum.  Therefore, the sample selection function becomes explicitly redshift-dependent, leading to a nonzero term $\partial\log\bar n/\partial z$ in the right-hand side of eq.\ \eqref{eq:dndz} which generates errors in eq.\ \eqref{eq:estimatebphi}. 

This source of error may be corrected by applying k-corrections and applying a revised selection based on k-corrected magnitudes, cut to remove the effects of scatter across boundaries. That is, we would subselect those targeted galaxies whose colours and magnitudes would pass the selection cut were the galaxy located at any redshift in the range being considered. Because spectra are measured for all targeted galaxies, the k-correction may be computed quite precisely. An absolute magnitude limited, k-corrected \cite{Hogg2002,Blanton2007} sample would have the required property, but this idea is applicable to more complicated selection criteria as well. In modern spectroscopic surveys like DESI \cite{DESI2024}, galaxies are targeted for spectroscopic observations using cuts on both their apparent magnitude and color \cite{Myers-DESI-targeting}, which complicates such a correction. However, it is still possible to select a subsample of galaxies that would have been targeted if they were placed at all redshifts in the range considered, and it should be possible to find a sub-sample with this property.

To be explicit, let us write the number density of galaxies at redshift $z$ and rest-frame spectrum $\s$ as $n(z,\s)$.  We select a subset of those galaxies as our sample, $\nsel(z) = \int n(z,\s) f(z,\s) d\s$, for some selection function $f$ that depends on the observed, redshifted spectrum, not the rest-frame spectrum, and therefore is a function of both \s\ and $z$.  We can construct a further subset of $\nsel(z)$ with a different selection function $F$ that depends only on \s, and not $z$, via  
\begin{equation} \label{eq:numsel}
	\nsel(z, z_1, z_2) = \int n(z,\s) f(z, \s) F(z_1,z_2,\s) d\s    ,
\end{equation}
where
\begin{equation} \label{eq:newsel}
	F(z_1,z_2,\s) = \prod_i f(z_i, \s) 
\end{equation}
and the product runs over all redshifts in the range $z_1 < z < z_2$.  The selection function $f$ is either 0 or 1, since galaxies are either targeted or they are not, and so $f^2 = f$.  Because of this, $f(z,\s) F(z_1,z_2,\s) = F(z_1,z_2,\s)$ for all redshifts in the range $z_1 < z < z_2$, and therefore eq.\ (\ref{eq:numsel}) reduces to
\begin{equation}
	\nsel(z, z_1, z_2) = \int n(z,\s) F(z_1,z_2,\s) d\s.
\end{equation}
We see that in this new sample, the only explicit dependence on $z$ is now only in $n(z,\s)$, and not in the selection function, as long as $z_1 < z < z_2$.  If $z$ falls outside the range $z_1$ to $z_2$, then the selection function still explicitly depends on $z$.  But within the range $z_1$ to $z_2$, the selection depends only on $z_1$ and $z_2$, not $z$, so we can once again estimate \bphi\ using $\partial\log\nsel/\partial z$ within that range.  Picking the optimal redshift range $z_1-z_2$ will involve balancing between two factors: using a wide enough range to measure redshift evolution, and a small enough range to avoid discarding too many galaxies with the modified selection function in eq.\ \eqref{eq:newsel}.  Note that the $z_1-z_2$ range does not have to encompass the entire observed redshift range, but rather we are free to subdivide the observed range into smaller chunks, over which we separately apply eq.\ \eqref{eq:newsel}.

\section{Summary} \label{sec:discussion}

We have considered whether we can measure \bphi\ directly by exploiting the change in $\sigma_8$ with redshift to estimate the derivative $\bphi = 2\partial\log\bar n/\partial\log\sigma_8$. This is only a good approximation for samples carefully selected such that the density of objects strongly depends on $\sigma_8$ and only weakly depends on other evolving quantities.  Using N-body simulations, 
we have shown that it is possible to choose samples of dark matter halos that satisfy this requirement, over a wide range of redshifts and masses.  It should be possible to construct similar samples of galaxies, if the halo occupation distribution of those galaxies can be determined using observations of their small-scale clustering.
For galaxy surveys we also have to consider the evolution of the selection function. We have proposed a method to mitigate this effect by subsampling, to construct a set of galaxies consisting of those that would have been selected at all redshifts within a particular sample or bin. Thus, a combination of abundance and small-scale clustering measurements may allow us to estimate \bphi, and thereby exploit large-scale clustering to probe \fnl\ and the physics of the earliest times in the universe.

\acknowledgments
We thank Selim Hotinli, Andrey Kravtsov, and Jamie Sullivan for helpful discussions.
Research at Perimeter Institute is supported in part by the Government of Canada through the Department of Innovation, Science and Economic Development Canada, and by the Province of Ontario through the Ministry of Colleges and Universities.
WP acknowledges support from the Natural Sciences
and Engineering Research Council of Canada (NSERC),
[funding reference number RGPIN-2019-03908] and from
the Canadian Space Agency. 
This research was enabled in part by support provided by Compute Ontario (computeontario.ca) and the Digital Research Alliance of Canada (alliancecan.ca).

\newcommand{\aap}{Astronomy \& Astrophysics}
\newcommand{\aaps}{Astronomy \& Astrophysics Supplement Series}
\newcommand{\aj}{Astronomical Journal}
\newcommand{\apj}{Astrophysical Journal}
\newcommand{\apjl}{Astrophysical Journal Letters}
\newcommand{\apjs}{Astrophysical Journal Supplement Series}
\newcommand{\jcap}{Journal of Cosmology and Astroparticle Physics}
\newcommand{\mnras}{Monthly Notices of the Royal Astronomical Society}
\newcommand{\prd}{Physical Review {\bf D}}
\newcommand{\physrep}{Physics Reports}
\bibliography{refs}

\end{document}